\documentclass{osa-article}

\journal{oe}

\articletype{Research Article}
\usepackage{soul} 
\soulregister\ref{7}
\soulregister\eqref{7}
\soulregister\cite{7}
\soulregister\onlinecite{7}

\begin{document}

\title{Introducing non-local correlations into laser speckles}

\author{Nicholas Bender,\authormark{1} Hasan Y{\i}lmaz,\authormark{1} Yaron Bromberg,\authormark{2} and Hui Cao \authormark{1*}}

\address{\authormark{1}Department of Applied Physics, Yale University, New Haven, CT 06520, USA\\
\authormark{2}Racah Institute of Physics, The Hebrew University of Jerusalem, Jerusalem, 91904, Israel}

\email{hui.cao@yale.edu} 



\begin{abstract}
Laser speckles have become a fundamental component of the modern optics-research toolbox. Not only are speckle patterns the basis of numerous imaging techniques, but also, they are employed to generate optical potentials for cold atoms and colloidal particles. The ability to manipulate a speckle pattern\textquotesingle s spatial intensity correlations, particularly long-range (non-local) ones, is essential in numerous applications. A typical fully-developed speckle pattern, however, only possesses short-ranged (local) intensity correlations which are determined by the spatial field correlations. Here we experimentally demonstrate and theoretically develop a general method for creating fully-developed speckles with strong non-local intensity correlations. The functional form of the spatial intensity correlations can be arbitrarily tailored without altering the field correlations. Our approach provides a versatile and utilitarian framework for enhancing and controlling non-local correlations in speckle patterns.
\end{abstract}

	\section*{Introduction}
	
	A bedrock principle of statistical physics is the Siegert equation, which relates the first and second-order correlation functions. It is the foundation of common techniques such as Hanbury-Brown Twiss interferometry\cite{brown1956correlation} and dynamic light scattering\cite{DLSbook}. Despite its general use, it is not universal. In quantum optics, for example, photon anti-bunching violates the Siegert equation. This violation has been widely explored in studies of non-classical light \cite{mandel1995optical}. For classical wave transport in mesoscopic systems, the violation of the Siegert equation is a hallmark of non-local correlations. Not only do non-local correlations reflect a proximity to Anderson localization, they are also responsible for universal conductance fluctuations \cite{ Akkermans}. Non-local correlations originating from crossed scattering paths in a disordered medium \cite{Cwilich, Mello2, Genack2, Lagendijk, Scheffold1, SebbahPRE00, CwilichPRE06, YamilovPRB08, Stone2, strudley2013mesoscopic},  however, are significantly weaker than the local correlations.  
	
	While previous mesoscopic physics studies have retrieved information about disordered systems from speckle correlations in scattered light \cite{DogariuReview}; this work explores the inverse process, namely, we design the scattering structure itself to obtain desired speckle intensity-correlations in the far field. In particular, we aim to enhance and manipulate non-local correlations by drastically, yet controllably, violating the Siegert relation so the local-correlations are unaffected. The simplest ``scattering'' structure -which can be facilely controlled- is a spatial light modulator (SLM). Although incident light is scattered once by the SLM, arbitrary correlations can be encoded among the SLM pixels. Such correlations can be significantly stronger and more versatile than correlations built among partial waves during the process of multiple scattering in a random medium. 
	
	We experimentally demonstrate that the speckle intensity correlation length can be augmented to significantly exceed the field correlation length: with non-local intensity correlations comparable in strength to the local intensity correlations. Furthermore, we show that it is possible to arbitrarily tailor the long-range intensity correlation function -for example making it anisotropic and oscillating- while keeping the field correlation function isotropic and untouched. Finally, a theoretical analysis reveals that the non-local intensity correlations in the far-field speckle patterns originate from high-order phase correlations encoded into the light field on the SLM plane.
	
	The ability to manipulate the intensity correlations of speckles has a plethora of potential applications. Speckle illumination has been used for computational imaging and compressive sensing. In this context, tailoring the speckle correlations would be essential for "smart" illumination of the target \cite{ DogariuOptica17}. In speckle-based fluorescence microscopy, the spatial intensity correlation function corresponds to the point spread function \cite{Bertolotti, mertz2011optical}, and thus customizing speckle correlations enables one to engineer the point spread function. Furthermore, laser speckle patterns with designed intensity correlations can be used as bespoke disordered optical-potentials in transport studies of cold atoms \cite{cold}, colloidal particles \cite{coll}, and active media \cite{active}.
	
	The spatial intensity correlation function is given by:
	\begin{equation} 
	C_{I}(\Delta {\bf r}) \equiv {\langle I({\bf r}) I({\bf r} + \Delta {\bf r})\rangle}/{\langle I({\bf r}) \rangle \langle I({\bf r} +\Delta {\bf r}) \rangle} -1 = C_{L} (\Delta {\bf r}) + C_{NL}(\Delta {\bf r}).
	\end{equation} 
	Here $C_{L} (\Delta {\bf r})$ is the local correlation function, and it is related to the field correlation function, $C_{E}(\Delta {\bf r}) \equiv {\langle E({\bf r}) E^{*}({\bf r} + \Delta {\bf r})\rangle}/{\langle |E({\bf r})|^2\rangle}$, by $C_{L}(\Delta {\bf r}) = |C_E(\Delta {\bf r})|^2$ \cite{Freund1001, GoodmanB, DaintyB, Mello2}. $C_{NL}(\Delta {\bf r})$ represents the non-local correlation \cite{BerkovitsPR94}, and it vanishes when the Siegert relation holds: $C_{I}(\Delta {\bf r}) = |C_E(\Delta {\bf r})|^2$. 
	
	Previous studies dedicated to altering speckle intensity correlations \cite{Asakura1, FractalSpec, ICorManip, Markovian, Guillon17, chriki2018rapid, waller2012phase, 2012_Fleischer_PRL,Battista1, Battista2} generally rely on the Siegert relation, and modulate the spatial field correlations. It is more challenging to violate the Siegert relation and control the intensity correlations without affecting the field correlations. Such a modification requires the field and intensity to fluctuate spatially on different length scales. Even in our recent demonstrations of speckle patterns with arbitrary intensity probability density functions, the field and intensity have the same correlation length \cite{Yaron, CSS}. Although speckled-speckles produced by double scattering have $C_{I}(\Delta r) \neq |C_{E}(\Delta r) |^2$, the difference $C_{I}(\Delta r) - |C_{E}(\Delta r) |^2$ representing the non-local intensity correlations $C_{NL}(\Delta r)$ is rather small \cite{ BarakatAO86, YoshimuraJOSAA92}. In the near-field zone of a scattering medium, the Sigert relation does not hold, but the speckles are not fully developed and have a low contrast \cite{JackA}. Here we develop a flexible yet robust method to introduce arbitrary non-local intensity correlations into fully-developed speckle patterns without altering the field correlations.
	
	\section*{Results}
	\subsection*{Enhanced non-local correlations}
	First, we demonstrate how to increase the intensity correlation length of the speckles in the far field of the SLM without altering the field correlation length. We begin by measuring a generic Rayleigh speckle pattern, Figs. 1(a) and 1(c), created in the far field with a random phase pattern is displayed upon the SLM. In this case, the speckle field obeys a circular Gaussian probability density function for the complex amplitudes, and possesses only short-range intensity correlations, $C_{I}(\Delta r) = |C_{E}(\Delta r) |^2$, as confirmed in Fig. 1(e). We then magnify the speckle intensity pattern numerically by a factor $\alpha$ to increase the intensity correlation length by the same factor. A nonlinear-optimization algorithm \cite{nLopt1, nLopt2} is used to determine a phase pattern -which upon application to the SLM- generates the enlarged speckle intensity pattern on the camera plane. To facilitate the convergence to a solution, we reduce the area we attempt to control -on the camera plane- to the central quarter of the region representing the Fourier transform of the phase modulating region of the SLM \cite{CSS}. Numerically we minimize the difference between the target intensity pattern and the intensity pattern obtained after applying the field transmission matrix (T-matrix, see Materials and Methods) to the SLM phase array. Since the SLM does not change the field amplitude, the spatial field correlation function in the Fourier plane remains identical to that of the unmagnified speckle pattern and therefore, so do the local intensity correlations $C_{L}(\Delta r) = |C_{E}(\Delta r) |^2$.
	
	After finding the appropriate two-dimensional (2D) SLM phase-patterns, we display them: recording the speckle patterns incident upon the CCD camera. Figure 1(b) and 1(d) present one demonstration of an ``enlarged Rayleigh'' speckle pattern. The intensity fluctuates on a length scale $\alpha = 2.5$ times longer than the Rayleigh pattern in Fig. 1(a). While the width of $C_{I}(\Delta r)$ is increased 2.5 times, $|C_{E}(\Delta r) |^2$ remains the same as the original Rayleigh speckles, as shown in Fig. 1(f). This means that the speckle field, more precisely, the phase of the field plotted in Fig. 1(d), fluctuates faster in space than the intensity. Still, the phase pattern is significantly modified relative to that of a Rayleigh speckle pattern such as in Fig. 1(c). It exhibits distinct topological features such as elongated equiphase lines, which can be see in Fig. 1(d). Nevertheless, these features are masked by the spatial averaging inherent to calculating the field correlation function. The dramatic difference between $C_{I}(\Delta r)$ and $|C_{E}(\Delta r) |^2$ demonstrates the profound non-local intensity correlations present in the speckle pattern. 
    
    \begin{figure}[hthb]
	\centering
	\includegraphics[width= 9cm]{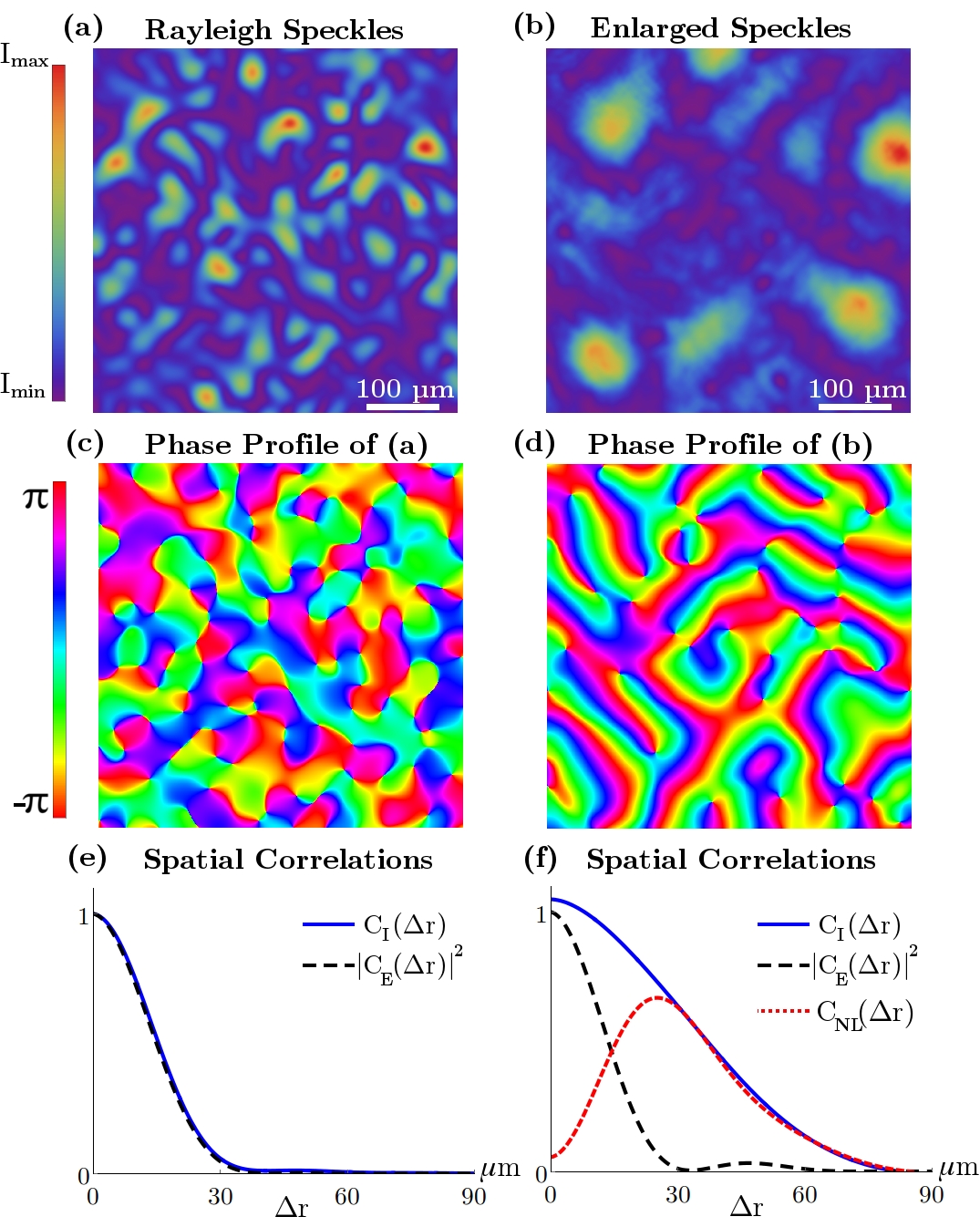}
	\caption{Enhancing non-local correlations in speckles. A Rayleigh speckle pattern (a,c) with $C_{I}(\Delta r) = |C_{E}(\Delta r)|^{2}$ (e), is compared to an "enlarged Rayleigh" speckle pattern (b,d) with $C_{I}(\Delta r)$ much broader than $|C_{E}(\Delta r)|^{2}$ (f). The non-local intensity correlations, $C_{NL}(\Delta r)$, have comparable strength to the local correlations, $C_{L}(\Delta r)=|C_{E}(\Delta r)|^{2}$, in (f). The correlation functions in (e,f) are obtained by averaging over 100 independent speckle patterns. Similar to the Rayleigh speckle pattern, the customized speckle field is fully developed with a uniform phase distribution between $0$ and $2 \pi$.}
	\label{Figure1}
	\end{figure}
\newpage

	Since the Rayleigh speckles are magnified by the same factor $\alpha = 2.5$ in both $x$ and $y$ directions, the intensity correlation functions, both  $C_{L}$ and $C_{NL}$, are isotropic and depend only on $\Delta r = |\Delta {\bf r}| = \sqrt{(\Delta x)^2 +(\Delta y)^2}$. Figure 1(f) compares $C_{NL}(\Delta r)$ to $C_{L}(\Delta r)$ and $C_{I}(\Delta r)$. Unlike $C_{L}$, $C_{NL}$ does not decay monotonically with $\Delta r$, instead it rises to its maximum when $C_{L}$ almost dies out, and subsequently $C_{NL}$ dominates the functional form of $C_{I}(\Delta r)$. The maximum value of $C_{NL}$ is comparable to that of $C_{L}$ at $\Delta r = 0$. In this example, the speckle intensity correlations become long-ranged but remain isotropic, namely, the correlation lengths are identical in both the $x$ and $y$ directions. We can easily make the correlations anisotropic, by setting the amplification factor in $x$ different from that in $y$, thereby tuning the intensity correlation lengths in $x$ and $y$ separately.

	\subsection*{Anisotropic long-range correlations}

	\begin{figure}[hthb]
		\centering
		\includegraphics[width= 9cm]{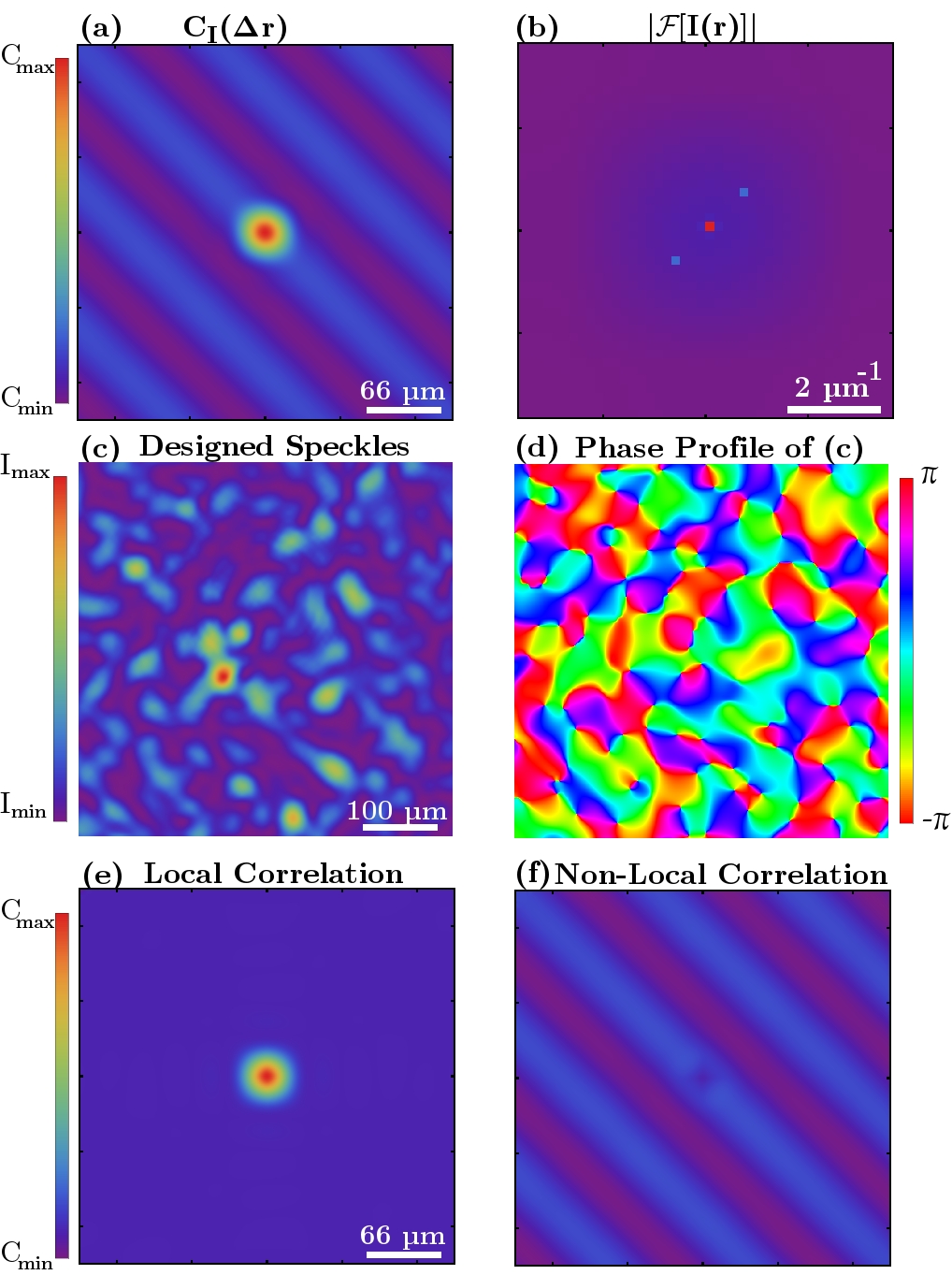}
		\caption{Creating speckle patterns with spatially oscillating, anisotropic long-range intensity correlations. The intensity correlation function $C_{I}(\Delta {\bf r})$ (a), determines the Fourier amplitude profile of $I({\bf r})$ (b). An experimentally generated speckle intensity-pattern $I({\bf r})$ (c) possessing the correlations given in (a), and the corresponding phase profile $\theta({\bf r})$ (d). $\theta$ is uniformly distributed between $0$ and $2 \pi$, confirming that the speckle pattern is fully developed. The local intensity correlation function $C_{L}(\Delta {\bf r})$ (e) has a maximum value of 1, while the non-local intensity correlation function $C_{L}(\Delta {\bf r})$ (f) has a maximum/minimum value of $\pm 0.1$. The correlation functions in (a,e,f) are obtained by averaging over 100  speckle patterns. The origins in (a,b,e,f) are located at the plots\textquotesingle\ centers.}
		\label{Figure2}
	\end{figure}
	
	Next, we demonstrate how to synthesize speckles with significantly more complex spatial intensity correlations.  Figure 2(a) shows $C_I(\Delta {\bf r})$ with an oscillating non-local correlation function $C_{NL}(\Delta {\bf r}) = (1/10) \cos[(\Delta {\bf x} + \Delta {\bf y})/{10}]$, where $x$ and $y$ are spatial coordinates. To generate speckles possessing such correlations, we first find speckle intensity patterns $I({\bf r})$ which adhere to the desired $C_I(\Delta {\bf r})$. Since the Fourier transform of $I({\bf r})$ is related to that of $C_I(\Delta {\bf r})$ by $\mathcal{F}[C_{I}(\Delta {\bf r})+1]=| \mathcal{F}[I({\bf r})]|^{2}$, $|\mathcal{F}[I({\bf r})]|$ is known. As plotted in Fig. 2(b), it is a sparse function.  We then solve for $I({\bf r})$ with a Gerchberg-Saxton algorithm. Starting with a Rayleigh speckle intensity pattern, $J({\bf r})$, we modify the amplitude of its Fourier components, such that $|\mathcal{F}[J({\bf r})]|$ is equal to $|\sqrt{\mathcal{F}[C_{I}({\bf \Delta r})+1]}|$, without altering the phase values. The inverse Fourier transform of the modified Fourier spectrum gives a complex valued function for $\tilde{J}({\bf r})$. Since intensity values must be positive real numbers, we ignore the phase values and set $\tilde{J}({\bf r}) = |\tilde{J}({\bf r})|$. Cyclical repetition of this process will eventually result in an intensity pattern which adheres to the desired correlation function. Starting with different initial Rayleigh speckle patterns will produce uncorrelated intensity patterns that satisfy the same $C_I(\Delta {\bf r})$. Using the nonlinear optimization algorithm discussed previously, we obtain the SLM phase patterns to create the desired intensity patterns on the camera. Figure 2(c) presents one such intensity pattern recorded experimentally. Its phase profile is predicted by the measured T-matrix and shown in Fig. 2(d). The local intensity correlation function $C_L(\Delta {\bf r})=|C_{E}(\Delta r)|^{2}$, shown in Fig. 2(e), remains isotropic and identical to that of the original Rayleigh speckles. However, the non-local correlation function $C_{NL}(\Delta {\bf r})$, plotted in Fig. 2(f), oscillates along the diagonal direction.

	\begin{figure}[htb]
		\centering
		\includegraphics[width= 9cm]{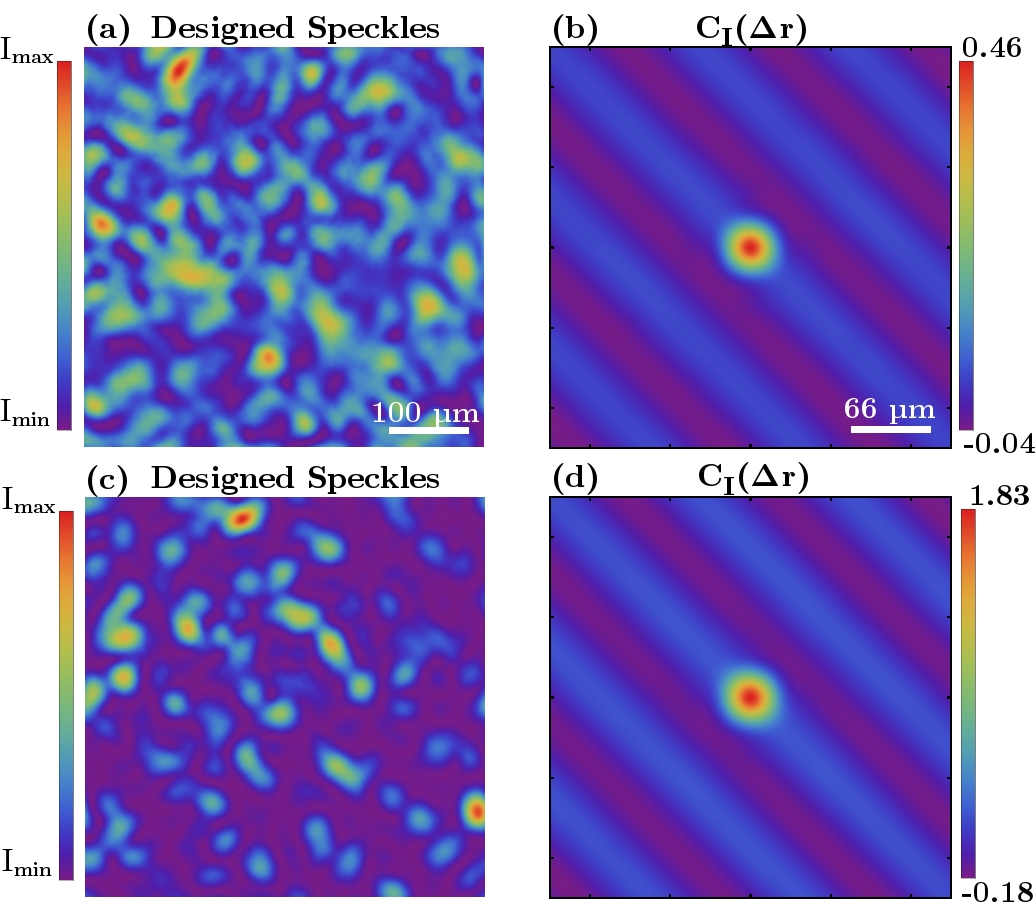}
		\caption{Tuning the speckle contrast independently from the spatial intensity correlation function. Two experimentally generated speckle patterns (a,c) with congruent intensity correlation functions (b,d). The intensity contrast is 0.68 in (a) and 1.35 in (c). The origin for (b) and (d) is located at the center of the plots.} \label{FigureS1}
	\end{figure}
	
	A useful feature of our method is its ability to vary the contrast of the speckle intensity without altering the functional form of the long-range intensity correlation function. For the example given in Fig. \ref{Figure2}(b), we can adjust the magnitude of the zeroth-order spatial frequency component, in order to change the constant background of the speckle intensity pattern in real space and thus modify the speckle contrast. Speckle patterns with identically shaped, i.e. congruent, $C_{NL}$ but different intensity contrasts are presented in Fig. \ref{FigureS1}(a) and \ref{FigureS1}(c). Given that the speckle contrast is directly related to the second moment of the intensity probability density function, this property illustrates the relative independence of the non-local correlations with respect to the intensity probability density function.

	\begin{figure*}[htb]
	\centering
	\includegraphics[width= \linewidth]{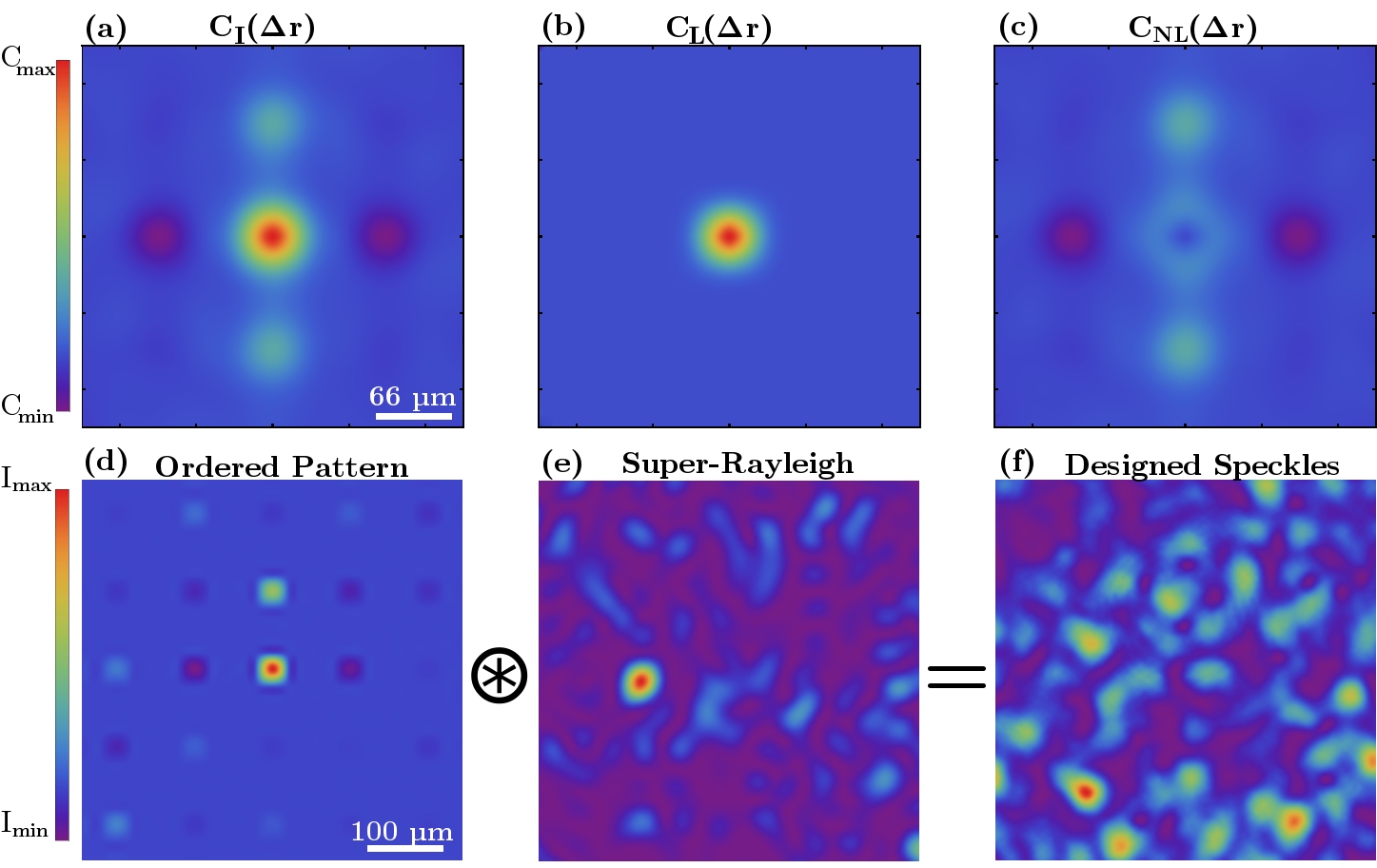}
	\caption{Introducing spatially simple anisotropic non-local correlations into speckles. The spatial intensity correlation function (a) is sparse. The local correlation function (b) has a maximum amplitude of 1, while the non-local correlation function (c) has a maximum/minimum amplitude of $\pm 0.2$. An ordered intensity pattern $g({\bf r})$ (d), produced by the Gerchberg Saxton algorithm, is convolved with super-Rayleigh speckle pattern $J({\bf r})$ (e) to generate a speckle intensity pattern $I({\bf r})$ (f) with the desired non-local correlations given in (c). The correlation functions in (a,b,c) are obtained by averaging over 100 speckle patterns and the origins are located at the plots\textquotesingle\ centers.}
	\label{Figure3}
	\end{figure*}
	
	Although the above method excels at generating speckle patterns when the desired non-local correlation function has sparse Fourier components, it fails to converge to a speckle pattern when the desired non-local correlation function is sparse in real space, such as the one shown in Fig. 4(a). While the correlations are positive at (0, 100 \textmu m) and (0, -100 \textmu m), they become negative at (100 \textmu m, 0) and (-100 \textmu m, 0). Rather than producing a random intensity pattern, the Gerchberg-Saxton algorithm converges to an ordered pattern, $g({\bf r})$ in Fig. 4(d), which adheres to the desired $C_I(\Delta {\bf r})$ in Fig. 4(a). To produce a speckle intensity pattern, we simply convolve $g({\bf r})$ with a speckle pattern without non-local correlations, such as $J({\bf r})$ in Fig. 4(e), and obtain $I({\bf r})= g({\bf r}) \circledast J({\bf r})$. This results in a speckle pattern with  $\mathcal{F}[ I({\bf r})] = \mathcal{F}[ J({\bf r})] \mathcal{F}[ g({\bf r})]$, and $\mathcal{F}[C_{I}(\Delta {\bf r})]\cong \mathcal{F}[ C_J({\bf r})] \mathcal{F}[ C_g({\bf r})]$. Since the local correlation length of the convolving speckle pattern is set by the diffraction limit, its correlation function can be approximated by a $\delta$ function \cite{Labeyrie}. Consequently, $\mathcal{F}[C_{I}(\Delta {\bf r})] \approx \mathcal{F}[C_g({\bf r})]$, and $I({\bf r})$ possesses the same intensity correlations as $g({\bf r})$. Once the target intensity-pattern $I({\bf r})$ is obtained, a corresponding speckle-pattern can be created experimentally using our nonlinear optimization algorithm: for example Fig. 4(f). Here the corresponding local and non-local intensity correlation functions are shown in Figs. 4(b) and 4(c). Just as before, one has the freedom to increase or decrease the speckle contrast of the target pattern, by convolving $g({\bf r})$ with either a super-Rayleigh or sub-Rayleigh speckle intensity pattern \cite{Yaron}.

	\subsection*{Origins of non-local correlations}	
	
	Next, we illustrate that the non-local intensity correlations introduced into the speckle patterns, $C_{NL}(\Delta {\bf r}) =C_{I}(\Delta {\bf r}) -|C_{E}(\Delta {\bf r})|^2$, originate from high-order correlations encoded in the phase patterns on the SLM. For simplicity,  we consider a 1D speckle field $E(r)$, of length $L$, and its spatial Fourier components, $\varepsilon(\rho)$, where $\rho$ corresponds to the spatial position on the SLM plane. 
	\begin{equation}
	E(r)  =   \frac{1}{\sqrt{L}} \sum^{L-1}_{\rho=0} \varepsilon(\rho) e^{i \frac{2 \pi}{L} r \rho}
	\end{equation}
	The spatial field correlation function is therefore given by:
	\begin{equation}
	C_{E}(\Delta r)= \frac{1}{L} \sum^{L-1}_{\rho=0} |\varepsilon(\rho)|^{2} e^{-i \frac{2 \pi}{L} \Delta r \, \rho}.
	\end{equation}
	Taking the absolute-value squared of this expression gives the local intensity correlation function $C_{L}(\Delta r)$: 
	\begin{equation}
	\label{Clocal}
	C_{L}(\Delta r) =  \frac{1}{L^2} \sum^{L-1}_{\rho_1,\rho_2=0} |\varepsilon(\rho_1)|^{2} |\varepsilon(\rho_2)|^{2}\nonumber e^{i \frac{2 \pi}{L} \Delta r (\rho_2-\rho_1)}.
	\end{equation}
	With an expression for the local correlations in hand, we turn to the spatial intensity correlations:
	\begin{equation}
	I(r)I(r+\Delta {\bf r}) = \frac{1}{L^{2}} \sum^{L-1}_{\rho_1,\rho_2, \rho_3,\rho_4=0} \varepsilon(\rho_1) \varepsilon^{*}(\rho_2) \varepsilon(\rho_3) \varepsilon^{*}(\rho_4) e^{i \frac{2 \pi}{L}{\big[} r(\rho_1-\rho_2) + (r+\Delta r)(\rho_3-\rho_4) {\big ]}}.
	\end{equation}
	Grouping the summation into four terms according to the number of different $\rho$'s summed over and spatial averaging gives:
	\begin{equation}
	\label{CI2}
	C_{I}(\Delta r)= C _{1}(\Delta r) + C_{2}(\Delta r) + C_{3}(\Delta r) +C_{4}(\Delta r)-1
	\end{equation}
	where:
	\begin{eqnarray}
	\label{C}
	C_{1}(\Delta r) & = & \frac{1}{L^{2}} \sum^{L-1}_{\rho_1 =0} |\varepsilon(\rho_1)|^{4}\nonumber \\
	C_{2}(\Delta r) & = & \frac{1}{L^{2}} \sum^{L-1}_{\substack{\rho_1,\rho_2=0\\\rho_1 \neq \rho_2}} |\varepsilon(\rho_1)|^{2} |\varepsilon(\rho_2)|^{2} (1 +e^{i \frac{2 \pi}{L} \Delta r(\rho_2-\rho_1)})\nonumber\\
	C_{3}(\Delta r) & = & \frac{2}{L^{2}} \Re {\Big [}\sum^{L-1}_{\substack{\rho_1,\rho_2=0\\\rho_1 \neq \rho_2}} \varepsilon(\rho_1)^{2} \varepsilon^{*}(\rho_2) \varepsilon^{*}(2\rho_1-\rho_2)e^{i \frac{2 \pi}{L} \Delta r(\rho_2-\rho_1)}{\Big ]}\nonumber \\
	C_{4}(\Delta r) & = & \frac{1}{L^{2}} \sum^{L-1}_{\substack{\rho_1,\rho_2, \rho_3=0\\\rho_1\neq \rho_2\neq \rho_3}} \varepsilon(\rho_1) \varepsilon^{*}(\rho_2) \varepsilon(\rho_3) \varepsilon^{*}(\rho_1-\rho_2+\rho_3)e^{i \frac{2 \pi}{L} \Delta r(\rho_2-\rho_1)}.
	\end{eqnarray}
	Since $C_1$ and $C_3$ are on the order of $1/L$, they are negligible for large $L$, and $C_I$ is dominated by $C_2$ and $C_4$:  
	\begin{equation}
	\label{C2}
	C_{I}(\Delta r) \simeq C_{2}(\Delta r) + C_{4}(\Delta r)-1.
	\end{equation}
	Comparing the expression of $C_{L}(\Delta r)$ to that of $C_{2}(\Delta r) -1$, we notice their difference scales as $1/L$. When $L$ is large, $C_{L}(\Delta r) \simeq C_{2}(\Delta r) -1$, and 
	\begin{equation}
	C_{I}(\Delta r) = C_{L}(\Delta r) +C_{4}(\Delta r)
	\end{equation}
	Therefore, the non-local correlation function $C_{NL}(\Delta r) \simeq C_{4}(\Delta r)$. The expression for $C_{4}(\Delta r)$ reveals that the non-local correlations originate from the fourth-order correlations between different Fourier components of the speckle fields.

\subsection*{Axial evolution of speckle correlations}

The tailored speckles will gradually lose the non-local correlations as they axially propagate away from the Fourier plane of the SLM. This can be understood in the Fresnel approximation, where the axial propagation of a field pattern adds a quadratic phase to its spatial Fourier spectrum \cite{GoodmanB}. Because the non-local intensity correlations result from high-order correlations encoded into the phases of the Fourier components, the phase parabola accompanying axial-propagation erodes away such correlations as the tailored speckles propagate axially (along $z$-axis), eventually only the local intensity correlations remain.  

	\begin{figure*}[htb]
	\centering
	\includegraphics[width= \linewidth]{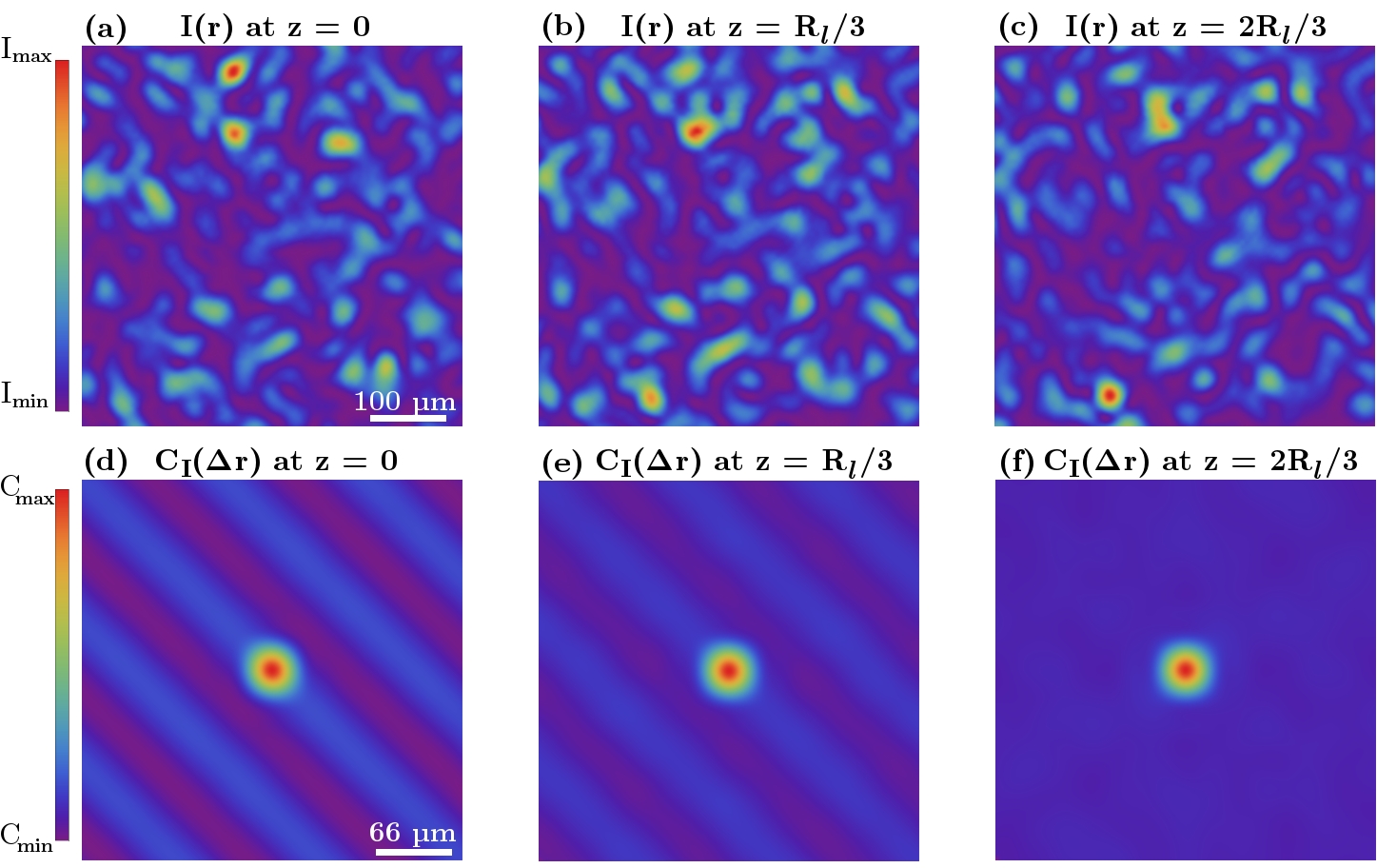}
	\caption{Axial evolution of a customized speckle pattern. An example customized speckle pattern (a), on the plane of customization $z=0$, is juxtaposed with it\textquotesingle s corresponding spatial intensity correlation function (d). The speckle pattern (b) and its intensity correlation function (e), after axially propagating to $z = R_{l}/3$, are presented. At this distance, the magnitude of the non-local correlations has reduced by half. The speckle pattern (c) and its intensity correlation function (f) are shown after further propagation to $z = 2 R_{l}/3$. At this point, the non-local correlations are completely erased and only the local correlations remain. The correlation functions in (d,e,f) are obtained by averaging over 100 different speckle patterns, and the origins are located at the center of the plots.}
	\label{Figure5}
	\end{figure*}
	
    In Fig. 5 we experimentally demonstrate the attenuation of non-local correlations in the customized speckles: as a function of axial propagation. Here we define $R_{l}$ as the axial correlation length of a speckle intensity pattern. It gives the average longitudinal size of a single speckle grain, and corresponds to the Rayleigh range. Figure 5(a) shows an example speckle pattern which is customized to have the oscillatory intensity-correlation function shown in Fig. 5(d) at $z = 0$. After the speckle pattern propagates a fraction of the Rayleigh-range $z=R_{l}/3$, Figs. 5(b), the non-local correlations attenuate to nearly half of their original magnitude, Fig. 5(d). Further propagation away from the plane of customization removes the remaining non-local correlations from the speckle pattern entirely: as can be seen in Figs. 5(c) and 5(f) for $z= 2R_{l}/3$. Beyond this point, the statistical properties of the customized speckles revert back to those of Rayleigh speckles.
	
	\section*{Discussion and conclusion}
	In conclusion, we presented a general approach for introducing strong non-local intensity-correlations into fully-developed speckle patterns using classical light in conjunction with a single scattering surface (SLM). By encoding fourth-order correlations into the phase of light reflected from the SLM, the second-order coherence function of the far-field speckles can be arbitrarily tailored without altering the respective first-order coherence function. Doing so, we drastically violate the Siegert relation: a fundamental principle in optical coherence theory. 
	
	Our method of encoding speckle correlations using the transmission matrix of an optical system is simple, yet versatile, and therefore can readily be incorporated into a broad range of optical experiments. For example, it would benefit studies of cold atom transport in correlated potentials \cite{ KuhnNJP07, PezzeNJP11, VincentPRL10}, because the spatial correlations of the speckled optical potentials could be arbitrarily customizable and re-configurable without the need for mechanical motion. Furthermore, our method can generate speckle patterns with desired correlations for illumination in compressive correlation imaging and stochastic optical sensing \cite{DogariuOptica17}. Since the spatial intensity correlation function determines photon coincidence counting rate, it is possible to create spatially correlated photon sources with tailored speckle patterns and engineer the coincidence counting rate for photon pairs as a function of their spatial separation.  
	
	Finally, it is worth mentioning the advantage of breaking the Siegert relation when controlling the intensity correlations of speckles. Methods relying on the Siegert relation modify the amplitude of light in the near field to control the spatial field correlations in the far field. Therefore, the total power of the far-field speckle pattern can be drastically reduced, which will degrade the sensitivity of imaging/sensing modalities using speckle illumination. Our method only requires phase modulation of the near field light, thus the total energy of the far-field speckle pattern is conserved.
	
	\section*{Materials and methods}
	Our experimental setup consists of a phase-only reflective SLM (Hamamatsu LCoS X10468) and a CCD camera (Allied Vision Prosilica GC660), which are juxtaposed at the front and back focal planes of a lens with focal length $f = 500$ mm. The SLM is uniformly illuminated by a linearly-polarized monochromatic laser beam at wavelength $\lambda = 642$ nm (Coherent OBIS). We only use the central part of the phase modulating region of the SLM, and partition it into a square array of $32 \times 32$ macro-pixels, each consisting of $16 \times 16$ pixels. The remaining illuminated pixels outside the central square diffract the laser beam away from the CCD camera via a phase grating. The SLM pixels can modulate the phase of the incident light between the values of $0$ and $2 \pi$ in steps of $2 \pi / 170$. While to a good approximation the field on the camera is a Fourier transform of the SLM field, we use an experimentally measured field-transmission matrix (T-matrix) to relate the light field on the SLM and camera planes in order to be more precise and general. For a given phase pattern displayed on the SLM, the differences between the speckle intensity patterns measured by the CCD camera and predicted by the field transmission matrix are negligible. Furthermore, using the measured field transmission matrix we numerically simulated the effects of the SLM\textquotesingle s dynamic range on the spatial correlations of the customized speckle patterns and found them to also be negligible. 

	\section*{Funding}
	US Office of Naval Research (N00014-13-1-0649). 


\bibliography{sample}

\end{document}